\documentclass[aps,pre,twocolumn,showpacs,amsfonts,amssymb,floatfix,superscriptaddress]{revtex4}
\usepackage[T1]{fontenc}
\usepackage[utf8]{inputenc}
\usepackage{float}
\usepackage{graphicx}
\usepackage{amssymb}
\usepackage{amsmath}
\usepackage{enumerate}

\makeatletter

\date{September 22, 2014} 

\makeatother

\begin{document}

\title{Comment on 'Anomalous diffusion induced by enhancement of memory'}

\author{Rüdiger Kürsten}
\affiliation{Institut für Theoretische Physik, Universität Leipzig, POB 100 920, D-04009 Leipzig, Germany}
\affiliation{International Max Planck Research School Mathematics in the Sciences, Inselstraße 22, D-04103 Leipzig, Germany}

\begin{abstract}
	In a recent paper \cite{Kim14} the author introduced and investigated a random walk model similar to a model introduced in \cite{ST04}. In these models the increment of the random walk depends on the complete past of the process. In this note I will point out that the models considered in \cite{ST04} and \cite{Kim14} can be mapped onto each other one to one. They can be defined on a common probability space and hence all expectation values of the model \cite{Kim14} with parameter $p$ are equal to the ones of \cite{ST04} with a corresponding parameter $\tilde{p}$.
\end{abstract}

\maketitle
In  section II of \cite{Kim14} a random walk model, that depends on the entire history of the process, is investigated. The author describes some differences to a similar model that was investigated in \cite{ST04} and is known as the elephant random walk (ERW). In both models there is a critical parameter $p_c$ such that for $p\le p_c$ the system shows normal diffusion, whereas for $p>p_c$ it shows super diffusion behavior. The author of \cite{Kim14} especially points out that the critical point in the model he considered is $p_c=1/2$ in contrast to \cite{ST04} where $p_c=3/4$. We will see in this comment that the two models are equivalent, when the corresponding parameters are identified in the correct way. The difference between the two models is only a relabeling of events. They can be defined on the same probability space such that all observables coincide for each realization. Hence also all expectation values coincide.

I will summarize the two discussed models. Both are one dimensional and discrete in space and time. In order to avoid confusion I will use the notation of \cite{ST04} for both models although the notation used in \cite{Kim14} is slightly different. In \cite{Kim14} the author considered the following random walk model (RWM). A random walker starts at $x=0$. In the first step he randomly goes either to $x=+1$ or to $x=-1$. The probability of going right or left in the first step is not precicesly specified. Let us introduce the parameter $q \in [0,1]$ as the probability to go right in the first step as in \cite{ST04}. The increment in the $(t+1)\text{-th}$ time step is denoted by $\sigma_{t} \in \{-1,1\}$ such that the position of the walker at time $t+1$ is
\begin{align}
	X_{t+1} = X_{t} + \sigma_{t}.
	\label{eq:update}
\end{align}
In all following steps starting from $t=1$ the increment $\sigma_t$ is chosen independently from the past with probability $1-p$, $p\in[0,1]$. In that case it will be $-1$ or $+1$ with equal probability. With probability $p$ the increment will be chosen from the past. A random number $t'$ is chosen with equal probability from $\left\{ 0, \cdots, t-1 \right\}$. Then $\sigma_{t} := \sigma_{t'}$.

The ERW is started at some position $X_0$. In this comment we will consider only $X_0=0$. The first step is chosen to be right with probability $q\in [0,1]$ and left with probability $1-q$. In all following steps a random number $t'$ is chosen uniformly from $\{ 0, \cdots, t-1\}$ as in the RWM. Then with probability $\tilde{p}\in [0,1]$ the increment $\sigma_t= \sigma_{t'}$ and with probability $1-\tilde{p}$, $\sigma_{t}=-\sigma_{t'}$.

Consider the ERW for $\tilde{p}\ge 1/2$. At time step $t$ assume we have chosen $t'$ uniformly from $\{0, \dots, t-1\}$. Let us consider the disjoint events $A$, $B$ and $C$ such that $\sigma_{t} = - \sigma_{t'}$ in case of $A$ and $\sigma_{t}= \sigma_{t'}$ in case of $B$ or $C$. These events shall have the probabilities $p(A)= p(B)= 1-\tilde{p}$ and $p(C)= 2\tilde{p}-1$. Then the probability that $\sigma_{t}=\sigma_{t'}$ is $\tilde{p}$ and the probability that $\sigma_t = - \sigma_{t'}$ is $1-\tilde{p}$ as in \cite{ST04}.\\
At this point we may ask our self why to consider the events $B$ and $C$ separately when they lead to the same increment. This will become clear when considering the events $C$ and $C'=A\cup B$. In case of event $C$, $\sigma_{t} = \sigma_{t'}$. And in case of the event $C'$, $\sigma_{t}=\sigma_{t'}$ or $\sigma_{t}=-\sigma_{t'}$, where both possibilities are equally likely. That means given $C'$, $\sigma_t$ is either plus or minus one with equal probability, no matter what the values of $t'$ and $\sigma_{t'}$ are. Hence we arrive at the RWM if we identify
\begin{align}
p=2\tilde{p}-1=p(C).
	\label{eq:paramid}
\end{align}
With this identity in mind we can compare some quantities calculated in \cite{Kim14} with the ones from \cite{ST04}. For example the recursion relation for the first moment given in Eq.~(11) of \cite{ST04} is
\begin{align}
	\langle x_{t+1}\rangle = \left( 1 + \frac{2\tilde{p}-1}{t} \right)\langle x_{t}\rangle.
	\label{eq:recfirstmomenttrimper}
\end{align}
In comparison Eq.~(7) of \cite{Kim14} is
\begin{align}
	\langle x_{t+1}\rangle = \left( 1+\frac{p}{t} \right)\langle x_{t}\rangle.
	\label{eq:recfirstmomentkim}
\end{align}
Similarly we find that the recursion relation for the second moment, the critical point and the Hurst exponent coincide when taking into account the identification Eq.~\eqref{eq:paramid}.

I want to remark that the ERW for $\tilde{p}\le 1/2$ can be identified with a similar model as the RWM, namely the following. The random walker starts at zero and in the first step randomly goes to the right or to the left as in RWM. In the next steps with probability $1-p$ he goes either left or right equally likely and with probability $p$ he remembers one of the previous steps and does the opposite of what he has done in the past, in contrast to RWM, where the walker does the same as in the past. In this sense the ERW is more general as it considers both cases at once.

As a last remark I want to point out that the process as it is considered in this comment is Markovian. This holds true if we consider the process that starts at $x=0$ or at some other fixed position $X_0$.
The reason for the Markovian nature of the process under consideration is the following. If the random walker decides to remember (or in the formulation of \cite{ST04} he alway remembers), what is essential for the outcome of the next increment is only the number of steps to the right and the number of steps to the left that have been made in the past. The random walker chooses to remember each increment of the past with the same probability. Therefore it is not important in which order the steps right and left have been performed. The number of steps to the right and the number of steps to the left that have been performed by the walker can be reconstructed uniquely from the position of the random walker and the total number of time steps that have been performed. Hence the distribution of $\sigma_t$, given $X_{t_1}, X_{t_2}, \dots, X_{t_n}$, with $t_1<t_2<\dots<t_n\le t$ is independent of $X_{t_1}, X_{t_2}, \dots, X_{t_{n-1}}$. 
If on the other hand the initial position of the process is randomly chosen, e.g. the process is started at $x=0$ with probability $1/2$ and it is started at $x=2$ with the same probability, the process is not Markovian. 

At this point it should have been made clear that the processes RWM and ERW for $\tilde{p}\ge \frac{1}{2}$ are equivalent. To make the argument more explicit I will introduce the probability spaces for the processes under consideration.

Let us consider the process until time $T\in \mathbb{N}$. Denote the set of realizations of ERW by $\tilde{\Omega}$. A realisation $\tilde{\omega} \in \tilde{\Omega}$ is the increment of the first step $\tilde{\sigma_0} \in \{-1, +1\}$ together with a pair of sequences $(\tilde{k})_{t}, (\tilde{S})_{t}$, $t\in\{1, \dots, T-1\}$. Here $\tilde{k}_t$ describes the past point in time the elephant is remembering at time $t$, therefore $\tilde{k}_t \in \left\{ 0, \dots, t-1 \right\}$. $\tilde{S}_t$ is one of the symbols $C$ or $C'$. When $\tilde{S}_t=C$ the elephant does the same as in the past. If $\tilde{S}_t=C'$ the elephant does the opposite of what he did in the past. The position $\tilde{X}_t(\tilde{\omega})$ of the elephant at time $t+1$ is determined by Eq.~\eqref{eq:update}, where the increments $\tilde{\sigma}_t$ are determined through the symbols $\tilde{S}_t$. If $\tilde{S}_t=C$ then $\tilde{\sigma}_t=\tilde{\sigma}_{k_{t}}$ and if $\tilde{S}_t=C'$ then $\tilde{\sigma}_t=-\tilde{\sigma}_{k_{t}}$.\\
Let us introduce the functions $\tilde{f}_t:(\tilde{k}_t, \tilde{S}_t) \rightarrow [0,1]$, $t\in \{1, \dots, T-1\}$ that are the probabilities that $\tilde{k}_t$ and $\tilde{S}_t$ have some particular value. We have $\tilde{f}_t(\tilde{k}_t=l, \tilde{S}_t=C)= \frac{1}{t}\tilde{p}$ and $\tilde{f}_t(\tilde{k}_t=l, \tilde{S}_t=C')=\frac{1}{t}(1-\tilde{p})$ for $l\in \{0, \dots, t-1\}$. Furthermore let us introduce the function $\tilde{f}_0 : \{+1, -1\}\rightarrow [0,1]$ that describes the probability of the first increment $\tilde{f}_0(+1)=q$, $\tilde{f}_0(-1)=1-q$. Then the probability of one realization is given by the product
\begin{align}
\tilde{P}_{\tilde{p}, q}(\tilde{\omega})= \tilde{f}_0(\tilde{\sigma}_0)\prod_{t=1}^{T-1}\tilde{f}_t(\tilde{k}_t, \tilde{S}_t),
	\label{eq:wskomega}
\end{align}
which defines a probability measure on $\mathcal{P}(\tilde{\Omega})$.
Hence the ERW $\tilde{X}_t$ is defined on the probability space $(\tilde{\Omega}, \mathcal{P}(\tilde{\Omega}), \tilde{P}_{\tilde{p},q})$.

Next I will introduce a probability space corresponding to the process RWM. Denote the set of realizations by $\Omega$. A realization $\omega \in \Omega$ is the increment of the first step $\sigma_0$ together with a sequence $(S)_t$, $t\in \left\{ 1, \dots, T-1 \right\}$. $S_t$ is either the symbol $A$, the symbol $B$ or a pair $(k, C)$, where $k\in \left\{ 0, \dots, t-1 \right\}$ and $C$ is a symbol. When $S_t=A$ then the increment is $\sigma_t(\omega)=+1$ independent of the past. When $S_t=B$ then $\sigma_t(\omega)=-1$ independent of the past. When $S_t=(k,C)$ then the increment is $\sigma_t(\omega)=\sigma_k$ that means the random walker remembers the increment at time $k$. Then the process $X_t(\omega)$ is given by Eq.~\eqref{eq:update}. Similar to the previous case we introduce the probability of $\sigma_0$ and $S_t$ as $f_0(+1)=q$, $f_0(-1)=1-q$, $f_t(S_t=A)= (1-p)/2=f_t(S_t=B)$, and $f_t(S_t=(k,C))=p/t$ for $t=1, \dots, T-1$.We observe that the probability that the increment is chosen independent from the past is $f_t(A)+f_t(B)=1-p$ as it is supposed to be for the RWM. The probability that the random walker is remembering is $\sum_{k=0}^{t-1}\frac{p}{t}= p$. As in the previous case the probability of a realization is given by the product
\begin{align}
	P_{p,q}(\omega)= f_0 \prod_{t=1}^{T-1}f_t(S_t),
	\label{eq:wskomega2}
\end{align}
which defines a probability measure on $\mathcal{P}(\Omega)$.
Hence the RWM $X_t$ is defined on the probability space $(\Omega, \mathcal{P}(\Omega), P_{p,q})$.

Now I will introduce a probability space that couples $(\tilde{\Omega}, \mathcal{P}(\tilde{\Omega}), \tilde{P}_{\tilde{p}, q})$ and $(\Omega, \mathcal{P}(\Omega), P_{p,q} )$. Therefore we consider the set of realizations $\hat{\Omega}$. A realization $\hat{\omega} \in \hat{\Omega}$ is the increment of the first step $\hat{\sigma_0}$ together with a pair of sequences $ (\hat{k})_t, (\hat{S})_t$, where $t\in \left\{ 1, \dots, T-1 \right\}$, $\hat{k}_t\in \left\{ 0, \dots, t-1 \right\}$ and $\hat{S}_t$ is one of the symbols $A$, $B$ or $C$.\\
We assign probabilities $\hat{f}_0$ on $\hat{\sigma}_0$ and $\hat{f}_t$ on $(\hat{k}_t, \hat{S}_t)$ as $\hat{f}_0(+1)=q$, $\hat{f}_0(-1)=1-q$, $\hat{f}_t(\hat{k}_t=l,\hat{S}_t=A)=\frac{1}{t}\frac{1-p}{2}=\hat{f}_t(\hat{k}_t=l, \hat{S}_t=B)$ and $\hat{f}_t(\hat{k}_t=l, \hat{S}_t=C)= \frac{p}{t}$ for $t\in \{1, \dots, T-1\}$, $l \in \{0, \dots, t-1\}$.\\
The product
\begin{align}
	\hat{P}_{p, q}(\hat{\omega}) = \hat{f}_0(\hat{\sigma}_0)\prod_{t=1}^{T-1}\hat{f}_t(\hat{k}_t, \hat{S}_t)
	\label{eq:wskomega3}
\end{align}
defines a probability measure on $\mathcal{P}(\hat{\Omega})$.

We can introduce a surjective map $\pi:\hat{\Omega} \rightarrow \Omega$, $(\hat{\sigma_0}, (\hat{k})_t, (\hat{S})_t) \mapsto ({\sigma_0}, ({S})_t)$ that keeps the increment of the first step $\sigma_0=\hat{\sigma_0}$, for each $t$ it maps $(\hat{k}_t=l, \hat{S}_t=A)$ to $S_t=A$, $(\hat{k}_t=l, \hat{S}_t=B)$ to $S_t=B$ and $(\hat{k}_t=l, \hat{S}_t=C)$ to $S_t=(l, C)$. One easily checks that
\begin{align}
	P_{p,q}(M) = \hat{P}_{p,q}(\pi^{-1}(M)) \qquad \forall M \subseteq \Omega.
	\label{eq:measureid1}
\end{align}
Hence the stochastic process 
\begin{align}
	\hat{X}_t(\hat{\omega}):= X_t(\pi(\hat{\omega}))
	\label{eq:xhat}
\end{align}
defined on the probability space $(\hat{\Omega}, \mathcal{P}(\hat{\Omega}), \hat{P}_{p, q})$ is equivalent to the stochastic process $X_t(\omega)$ defined on the probability space $(\Omega, \mathcal{P}(\Omega), P_{p, q})$.\\
One the other hand we can define a surjective map $\tilde{\pi}: \hat{\Omega} \rightarrow \tilde{\Omega}$, $(\hat{\sigma_0}, (\hat{k})_t, (\hat{S})_t) \mapsto (\tilde{\sigma_0}, (\tilde{k})_t, (\tilde{S})_t)$ that keeps the increment of the first step $\hat{\sigma_0}=\tilde{\sigma_0}$. It also keeps the sequence of times to remember $(\hat{k})_t=(\tilde{k})_t$. $\tilde{S}_t$ is mapped in the following way. If $\sigma_t(\pi(\hat{\omega}))=\sigma_{\hat{k}_t}(\pi(\hat{\omega}))$ then $\tilde{S}_t=C$ and otherwise $\tilde{S}_t=C'$.
From the definition of the maps $\pi$ and $\tilde{\pi}$ we see that $\hat{X}_t$ defined by Eq.~\eqref{eq:xhat} satisfies
\begin{align}
	\hat{X}_t(\hat{\omega})= \tilde{X}_t(\tilde{\pi}(\hat{\omega})) \qquad \forall \hat{\omega}\in \hat{\Omega}.
	\label{eq:xhat2}
\end{align}
One finds that the probability measures $\hat{P}_{p,q}$ and $\tilde{P}_{\tilde{p}, q}$ are related as
\begin{align}
	\tilde{P}_{\tilde{p}=\frac{1}{2}(p+1),q}(M) = \hat{P}_{p,q}(\pi^{-1}(M)) \qquad \forall M \subseteq \tilde{\Omega}.
	\label{eq:measureid2}
\end{align}
From the last two equations we conclude that the stochastic process $\hat{X}_t$ defined on the probability space $(\hat{\Omega}, \mathcal{P}(\hat{\Omega}), \hat{P}_{p,q})$ is equivalent to $\tilde{X}_t$ defined on $(\tilde{\Omega}, \mathcal{P}(\tilde{\Omega}), \tilde{P}_{\tilde{p}=\frac{1}{2}(p+1), q})$. Therefore also the processes $X_t$ and $\tilde{X}_t$ are equivalent. If we consider ERW and RWM to be defined on the same probability space $(\hat{\Omega}, \mathcal{P}(\hat{\Omega}), \hat{P}_{p,q} )$ via Eqs.~\eqref{eq:xhat},\eqref{eq:xhat2}, we observe that
\begin{align}
	\tilde{X}_t(\hat{\omega}) = X_t(\hat{\omega}) \qquad \forall \hat{\omega}\in \hat{\Omega},
	\label{eq:identity}
\end{align}
that is ERW and RWM are identical.

\end{document}